\documentstyle[preprint,aps]{revtex}
\begin{document}
\draft
\tighten
%
\title{Matter radii of light halo nuclei}

\author{J.S. Al-Khalili and J.A. Tostevin}
\address{Department of Physics, University of Surrey, Guildford,\\ Surrey, GU2
5XH, U.K.}
\date{\today}

\maketitle
 
\begin{abstract}
We re-examine the matter radii of diffuse halo nuclei, as deduced from reaction
cross section measurements at high energy. Careful consideration is given to
the intrinsic few-body structure of these projectiles and the adiabatic nature
of the projectile-target interaction. Using $^{11}$Li, $^{11}$Be and $^{8}$B as
examples we show that data require significantly larger matter radii than
previously reported. The revised value for $^{11}$Li of 3.55 fm is consistent
with three-body models with significant $1s$-intruder state components, which
reproduce experimental $^{9}$Li momentum distributions following $^{11}$Li
breakup, but were hitherto thought to be at variance with cross section data.
\end{abstract}

\pacs{PACS numbers: 24.10.Ht, 24.50.+g, 25.10.+s, 21.10.Gv, 11.80.Fv }


Reaction cross section measurements at energies of several hundred MeV/nucleon
have been used to study the radial extent of matter densities of short lived
exotic nuclei produced by fragmentation \cite{tanihata,Tan1}. Extensive tables
of deduced radii are now available in the literature, e.g.  \cite{Tan88b}.
Glauber theoretical methods \cite{Glaub,Czyz} have been the basis for these
assignments, and in particular the approximation \cite{Tan88b,Czyz} in which it
is assumed that the projectile and target nuclei present static density
distributions \cite{aside} whose geometric overlap determines the reaction
cross section.  To high accuracy, the deduced rms radii are found to be
essentially independent of the details of the projectile density distributions
assumed, e.g.  \cite{Tan88b,chul}. The accuracy of such deduced root mean
square (rms) radii is of considerable importance since they are routinely used
as empirical measures in constructing, constraining and assessing theoretical
models of halo structures for use in the interpretation of data.

At the heart of the static density model is the neglect of correlations between
the projectile (and target) constituents, each projectile nucleon being assumed
to carry the same single particle density \cite{Czyz}.  This assumption would
appear to work well for spatially localized nuclei such as $^{12}$C
\cite{Kox}.  For weakly bound systems such as halo nuclei, however, the
intrinsic few-body nature or granularity of the projectiles imply strong
spatial correlations between the valence nucleons and the more localized core.
At incident energies of order 800 MeV/nucleon one must also consider the
relevant timescales for a significant motion of these valence particles inside
the projectile and that for the passage of the same particle through the target
interaction region. In breakup studies narrow momentum widths are associated
with these valence particles which have characteristic kinetic energies of
order 10--40 MeV within the projectile \cite{Tommo}. For this reason reaction
models \cite{Og,att} make an adiabatic approximation, freezing the position
coordinates of the few-body projectile constituents during the interaction.
Physical observables are then obtained by suitably averaging the resulting
position dependent reaction amplitudes over the relevant position probability
distributions of these constituents.

This few-body picture suggests a quite different description of the
projectile-target interaction and formulation of the reaction cross section.
Consider for example $^{11}$Li as a pair of neutrons bound to a $^9$Li core.
For an impact parameter $b$ of the $^{11}$Li center of mass, Figure 1, such
that the projectile static density (shaded circle) overlaps the target, many
spatial configurations of the constituent bodies will not overlap the target.
The expectation is that the valence nucleon (large $b$) contribution to the
reaction cross section will be reduced or, alternatively, that the collision
will appear more transparent than otherwise expected.  Nishioka and Johnson
\cite{nj} investigated related adiabatic effects on light-ion composite
projectile (d, t, $^{3}$He and $\alpha$) cross sections in the energy range
$100\leq E\leq 350$ MeV/nucleon. The effects were very significant for the
extended deuteron but small for the $\alpha$ particle. Estimates of the
accuracy of the static density model for $^{11}$Li were considered earlier by
Takigawa {\em et al.} \cite{Tak} for a simplified two-body (di-neutron) halo
density and at lower energies, where the adiabatic limit is expected to be less
reliable. They demonstrated clearly the convergence of the two-body and static
density models in the limit of tight valence nucleon binding.  They concluded
that static density calculations would indeed overestimate reaction cross
sections, however the model used was too crude to allow a quantitative
discussion. The overestimation of cross sections in the static density model
was also recognised previously by Chulkov {\em et al.} \cite{chul}.

In this Letter we examine the quantitative implications of the few-body
adiabatic description for deduced matter radii of halo nuclei. We take as
examples two-body, $^{8}$B and $^{11}$Be, and three-body, $^{11}$Li, systems
for which cross section data are available for each composite and core
($^{7}$Be, $^{10}$Be, and $^{9}$Li) nucleus on a $^{12}$C target at 800
MeV/nucleon. Data are also available for the nucleon--$^{12}$C system, so that
all projectile constituent-target subsystems can be interrogated and compared
with experiment. In common with the analysis of Takigawa {\em et al.}
\cite{Tak} we will apply the static density approximation to the spatially
localized core-target and valence nucleon-target subsystems.  Additionally, the
adiabatic (frozen coordinate) treatment of these constituents allows us to
study carefully the implications of a realistic treatment of the two- and
three-body nature of the projectile wavefunctions on calculated cross
sections.
 

In Glauber theory \cite{Glaub} the reaction cross section for projectile $P$ is 
\begin{eqnarray}
\sigma_R (P) = 2\pi \int_{0}^{\infty}db\,b\left[1-T_P(b) \right]~~,\label{olxs}
\end{eqnarray}
where $T_P(b)$, the squared modulus of the Glauber $S$-matrix, is the
transparency of the collision at impact parameter $b$ of the projectile center
of mass (cm). In the static density limit 
\begin{eqnarray}
T_P^{\text{SD}}(b)=\exp \left[ -\bar{\sigma}_{\rm NN}^{PT} \int d^2{\bf
x}\,\rho_P^{(z)}(\vert{\bf x}\vert) \,\rho_T^{(z)}(\vert {\bf b}-{\bf x} \vert
) \right]~~ ,\label{two}
\end{eqnarray}
where $\bar{\sigma}_{\rm NN}^{PT}$ is the free nucleon-nucleon (NN) cross
section, at the relevant energy, appropriate for the projectile and target
\cite{charagi} with densities $\rho_P$ and $\rho_T$, and the
\begin{eqnarray}
\rho_i^{(z)}(b)=\int_{-\infty}^{\infty} dz\ \rho_i(\sqrt{b^2+z^2})~~
\end{eqnarray}
are the $z$-integrated densities or thickness functions. Here only the
projectile ground state density enters the calculation and few-body
correlations, the granular nature of the projectile, does not enter
explicitly.

In the few-body adiabatic limit, the transparency function is \cite{Jap},
\begin{eqnarray}
T_P^{\text{AD}}(b) = \vert \langle \Phi_0^n \vert S_C(b_C)S_v (b_v)
\vert \Phi_0^n \rangle \vert ^2 ~~,\label{adtran}
\end{eqnarray}
where $|\Phi_0^n \rangle$ is the wavefunction for the relative motion of the
$n$-constituent bodies in the projectile ground state, the bra-ket denoting
integration over these internal coordinates. For a
two-body (one valence nucleon+core) projectile the core-target and valence
nucleon-target $S$-matrices, in the static density limit, are 
\begin{eqnarray}
S_C(b_C)=\left[T_C^{\text{SD}}(b_C)\right]^{1/2} ~~,\qquad S_v(b_v) \equiv
S_1(b_1)=\left[T_N^{\text{SD}}(b_1)\right]^{1/2} ~~,
\end{eqnarray}
with $b_C$ the impact parameter of the core and $T_N^{\text{SD}}$ the analogue
of Eq.\ (\ref{two}) for the nucleon. For a three-body (two valence
nucleon+core) system, then of course
\begin{eqnarray}
S_v(b_v) \equiv S_1(b_1) S_2(b_2) ~~,\label{gamgam}
\end{eqnarray}
where the co-ordinates, in the plane perpendicular to the beam direction, are
shown in Figure 2. Eqs.\ (\ref{olxs}) through (\ref{gamgam}) are calculated
exactly in the following for realistic two- and three-body wavefunctions
$|\Phi_0^n \rangle$.  The explicit forms of the three-body wavefunction for
$^{11}$Li are given in \cite{att}.


We apply the formalism above to calculate reaction cross sections in the static
density and adiabatic limits for the one- and two-neutron halo nuclei $^{11}$Be
and $^{11}$Li, and the one-proton halo nucleus candidate $^8$B, all on a
$^{12}$C target at 800 MeV/nucleon. The choice of energy and target was
dictated by our wish to connect cross sections for all binary sub-systems with
experiment.

For all three incident nuclei, the static density calculations of the
projectile-target ($T_P^{\text{SD}}$), core-target ($S_C$) and valence
particle-target ($S_v$) sub-systems use the prescription for $\bar{\sigma}_{\rm
NN}^{iT}$ $(i=P,C,N$) of Charagi and Gupta \cite{charagi}. A Gaussian matter
distribution is assumed for $^{12}$C in all cases with rms matter radius
$\langle r^2\rangle^{1/2}_{12}=2.32$ fm \cite{Tan88b}. With these inputs, and
assuming Gaussian matter distributions for the core nuclei with radii $\langle
r^2 \rangle^{1/2}_9= 2.30$ fm, $\langle r^2\rangle^{1/2}_{10}=2.28$ fm and
$\langle r^2\rangle^{1/2}_{7}= 2.31$ fm, we calculate reaction cross sections
for the core-target subsystems $\sigma_R (^9$Li$)=796$ ($796\pm~\!\!6$) mb,
$\sigma_R (^{10}$Be$)=813$ ($813\pm 10$) mb and $\sigma_R (^7$Be$)=738$
($738\pm 9$) mb. The empirical values, in parentheses, are taken from
\cite{Tan1}. The deduced core radii agree with those of \cite{Tan88b} within
error bars.  The calculated nucleon-$^{12}$C cross section at 800 MeV is
$\sigma_R (N)=231$ mb which also agrees with experiment \cite{mmm} within
quoted errors. Thus each projectile constituent-target input to the few-body
calculations, the $S_C$ and $S_v$, is consistent with independent empirical
data for that binary system.

Part (a) of Figure 3 shows the results of static density and adiabatic
calculations for the $^{11}$Li+$^{12}$C system for a number of theoretical
three-body wavefunctions of $^{11}$Li. We show the calculated cross sections
versus the matter rms radius calculated from the wavefunction models.  The
horizontal band shows the experimental interaction cross section datum $\sigma
(^{11}$Li$)=1060 \pm 10$ mb \cite{Kob} and the vertical dashed line the
previously quoted matter radius $\langle r^2\rangle^{1/2}_{11}=3.10 \pm 0.17$
fm \cite{Tan88b}.

The (upper) open symbols are the results of the static density model and the
(lower) full symbols those of the adiabatic calculations for each wavefunction
model. The reduction in the calculated cross sections, or increased
transparency of the projectile in the latter case, is immediately evident. From
left to right the diamond symbols correspond to the P0 through P4 intruder
$s$-wave (Faddeev) model wavefunctions of Thompson and Zhukov \cite{taZ}, with
increasing rms radius. The extreme right hand point is a continuation of these
model wavefunctions (P5) with a $1s$-state scattering length of $-44$ fm and
80\% $(1s_{1/2})^2$ probability. The upright and inverted triangles are
calculations using the L6A pairing model wavefunction \cite{BT}, which in the
static density picture fits the published radial value, and the weak binding
potential $0s$-wave intruder wavefunction (G1 of \cite{taZ}) inspired by the
work of Johannsen, Jensen and Hansen \cite{jjh}.  The straight lines through
these model points are to guide the eye.

The results of these calculations are indeed dramatic. Whereas static density
calculations suggest a matter rms radius of order 3.1 fm, as reported
previously, a correct treatment of the $^{11}$Li three-body character now
suggests the halo is very much more extended and that $\langle r^2
\rangle^{1/2}_{11}= 3.55 \pm 0.10$ fm, firmly in the middle of the range of
values generated by intruder state models which successfully reproduce
empirical breakup momentum distributions \cite{taZ}.

Part (b) of Figure 3 shows the results of similar calculations but for the
one-neutron halo system $^{11}$Be. Again the horizontal band
shows the experimental cross section datum $\sigma (^{11}$Be$)=942 \pm 8$ mb
\cite{Tan1} and the vertical dashed line the previously reported
rms matter radius $\langle r^2\rangle^{1/2}_{11}=2.71 \pm 0.05$ fm
\cite{Tan88b}.  The results are qualitatively very similar to those of the
three-body $^{11}$Li case.  The angled dashed line shows the static density
calculations and the angled solid line and full symbols the adiabatic model
results. In this case these lines connect a large number of calculations using
simple two-body ($1s_{1/2}$) cluster wavefunctions for $^{11}$Be using binding
potentials with a range of geometries and with depth adjusted to reproduce the
single neutron separation energy 0.503 MeV. The solid symbols are
the results of adiabatic calculations for $^{11}$Be wavefunctions \cite{fn}
which include the effects of core ($^{10}$Be) deformation and excitation. The
wavefunction with rms radius 2.92 fm, whose calculated cross section lies
within experimental error bars, best reproduces the excited state spectrum of
$^{11}$Be. These wavefunctions generate cross sections which follow precisely
the trend of the inert core calculations and suggest a revised matter rms
radius of $\langle r^2\rangle^{1/2}_{11} = 2.90 \pm 0.05$ fm.

Finally, in part (c) of Figure 3 we consider the one proton-halo nucleus
candidate $^8$B. The previously reported value of $\langle r^2
\rangle^{1/2}_{8} = 2.39 \pm 0.04$ fm \cite{Tan88b} was very close to that for
$^7$Be, $\langle r^2\rangle^{1/2}_{7}=2.33 \pm 0.02$ fm \cite{Tan88b}
suggesting, in spite of the very small proton separation energy (0.137 MeV)
that the last proton had rather limited extension. The experimental cross
section for $^8$B has recently been revised to $\sigma (^8$B$)=798 \pm 6$
mb \cite{tanpc} and is shown by the horizontal band on the figure. Using the
static density model and a Gaussian density, in the manner of \cite{Tan88b}, we
obtain a revised static density estimate of $\langle r^2\rangle^{1/2}_{8}=2.42
\pm 0.03$ fm, shown by the vertical dashed line. The angled dashed and solid
lines are the results of static density and adiabatic model calculations for a
large number of two-body ($0p_{3/2}$) cluster wavefunctions for $^{8}$B based
on Woods-Saxon potential geometries. The diamonds use wavefunctions based on
the often used cosh form cluster model interaction \cite{bbb} and lie on the
same lines. Although the differences between the model calculations are smaller
than in the neutron halo cases, they remain very significant and suggest the
rms radius of $^{8}$B should be revised to $\langle r^2\rangle^{1/2}_{8}=2.50
\pm 0.04$ fm, indicating quite significant extension of the last proton
distribution beyond that of the core.


In summary, we have reanalyzed experimental data of reaction cross sections for
$^{11}$Be, $^{11}$Li and $^8$B projectiles on a $^{12}$C target at 800
MeV/nucleon using an adiabatic treatment of the internal coordinates of the
two- and three-body projectiles. We verify that all binary channel inputs to
the adiabatic model are consistent with the available experimental data for
these independent systems.  The granular structure of the projectiles implied
by realistic few-body wavefunctions is shown to reduce considerably the
calculated reaction cross sections and increase significantly the values of
matter rms radii deduced from data when compared to static density estimates.

We deduce matter rms radii for $^{11}$Li, $^{11}$Be and $^8$B of $3.55 \pm
0.10$ fm, $2.90 \pm 0.05$ fm and $2.50 \pm 0.04$ fm, respectively, representing
increases of 14.5\%, 7\% and 4.6\% over previously tabulated values. Our
revised radius for $^{11}$Li is now consistent with theoretical three-body
models with a significant $1s$-wave intruder state component, which reproduce
breakup momentum distributions, but were hitherto thought to be at variance
with cross section data.  Our revised radius for $^{11}$Be is also consistent
with two-body models which include core excitation and reorientation effects.

The increased transparency of the few-body structures presented here is quite
general, has implications for the deduced radii of all such exotic systems, and
suggests that a careful re-examination of all such data is necessary. The
particular importance of these effects in extended three-body halo systems is
exciting. In the case of $^{11}$Li we show this to be of importance in
elucidating their structure and in bringing consistency between calculations
and data for both breakup momentum distributions and reaction cross sections.

\acknowledgments 
The financial support of the United Kingdom Engineering and Physical Sciences
Research Council (EPSRC) in the form of Grants GR/J95867 and GR/K33026 is
gratefully acknowledged. We would like to thank Drs Filomena Nunes and Ian
Thompson for providing two- and three-body wavefunctions for the $^{11}$Be and
$^{11}$Li systems and Mr Matthew Bush for providing elements of the static
density Glauber model code used here. We thank Professor Ron Johnson for very
useful comments and discussions resulting from an earlier draft of this
letter.

\begin{figure}
\caption{Schematic representation of the static density (shaded circle) and
few-body adiabatic (frozen coordinate) treatments of the three-body projectile
(P)-target (T) collision at impact parameter $b$. In the spatial configuration
drawn the few-body projectile does not overlap the target.}
\end{figure}

\begin{figure}
\caption{Definition of position coordinates, in the plane perpendicular to the
beam direction, in the case of a three-body (two valence nucleon+core)
projectile.}
\end{figure}

\begin{figure}
\caption{Calculated static density and few-body adiabatic reaction cross
sections at 800 MeV/nucleon incident energy as a function of projectile rms
matter radius, for a $^{12}$C target. Parts (a), (b) and (c) of the figure are
for $^{11}$Li, $^{11}$Be and $^{8}$B projectiles, respectively. Details are
given in the text.  }
\end{figure}


\begin{references}
\bibitem{aside} The simplified Glauber theory used \protect\cite{Tan88b,Czyz}
is often called the optical limit, a term used in a number of contexts in the
literature. We refer to this approximation as the static density limit in the
present work.
\end{references}
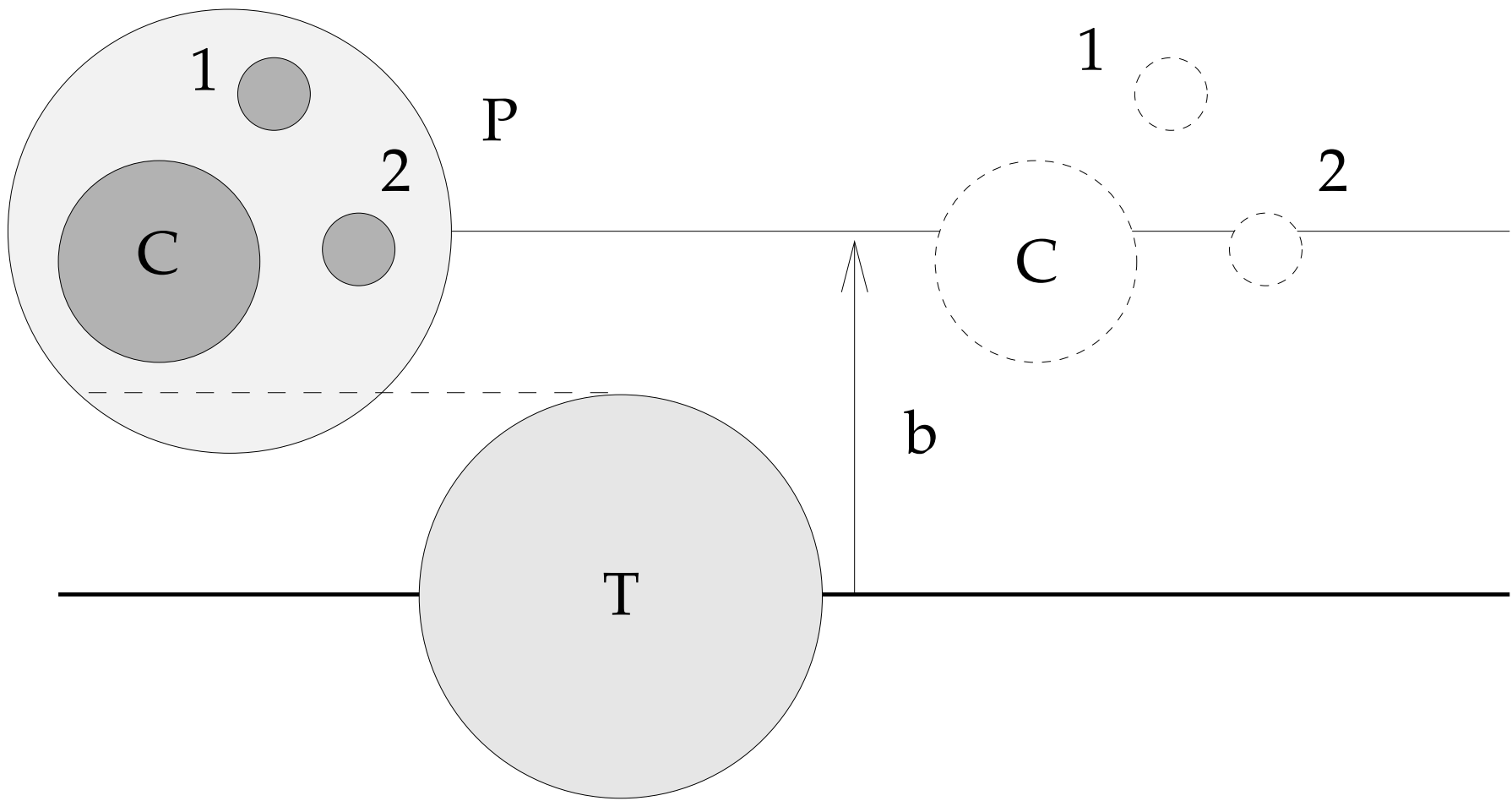
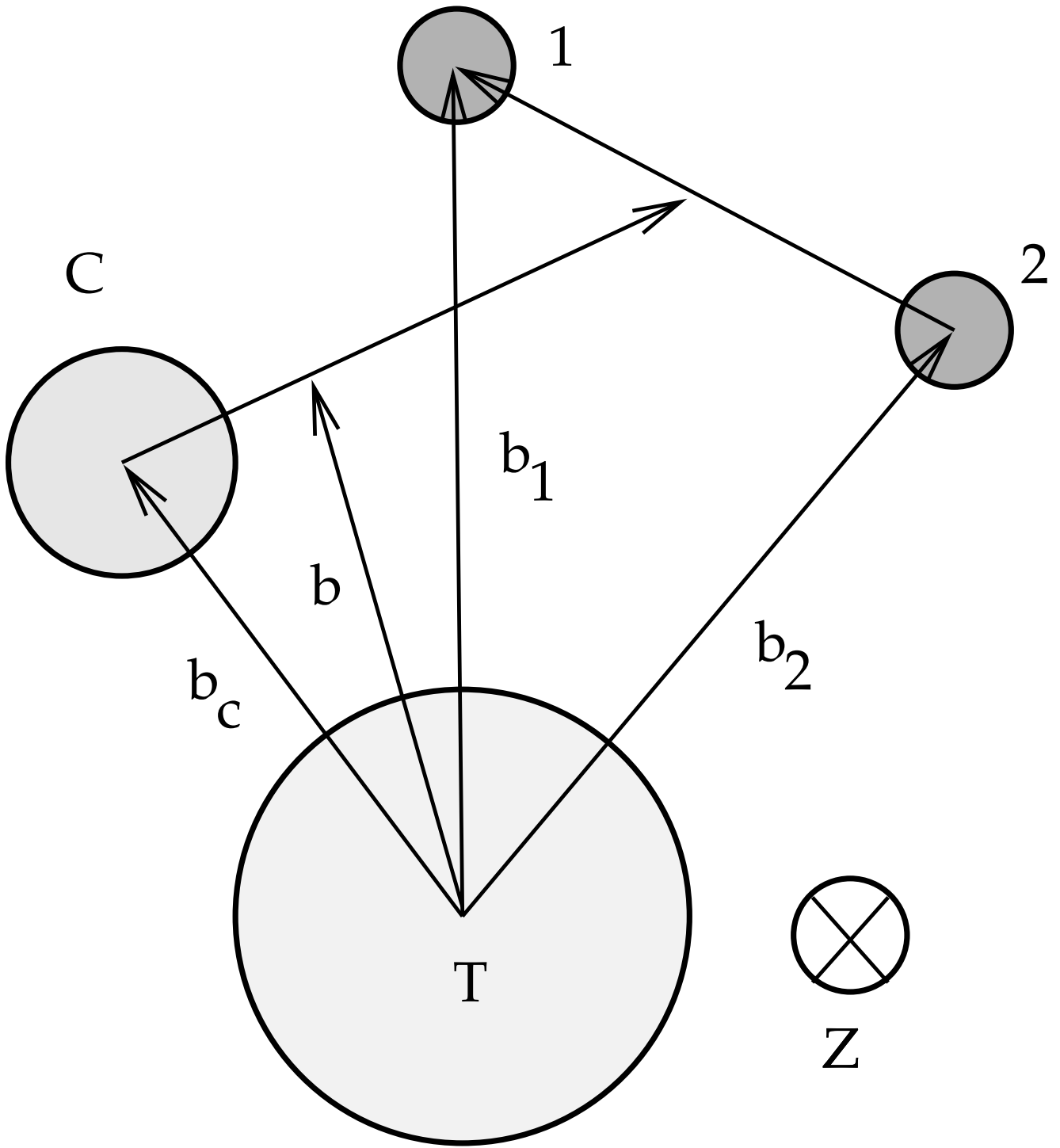
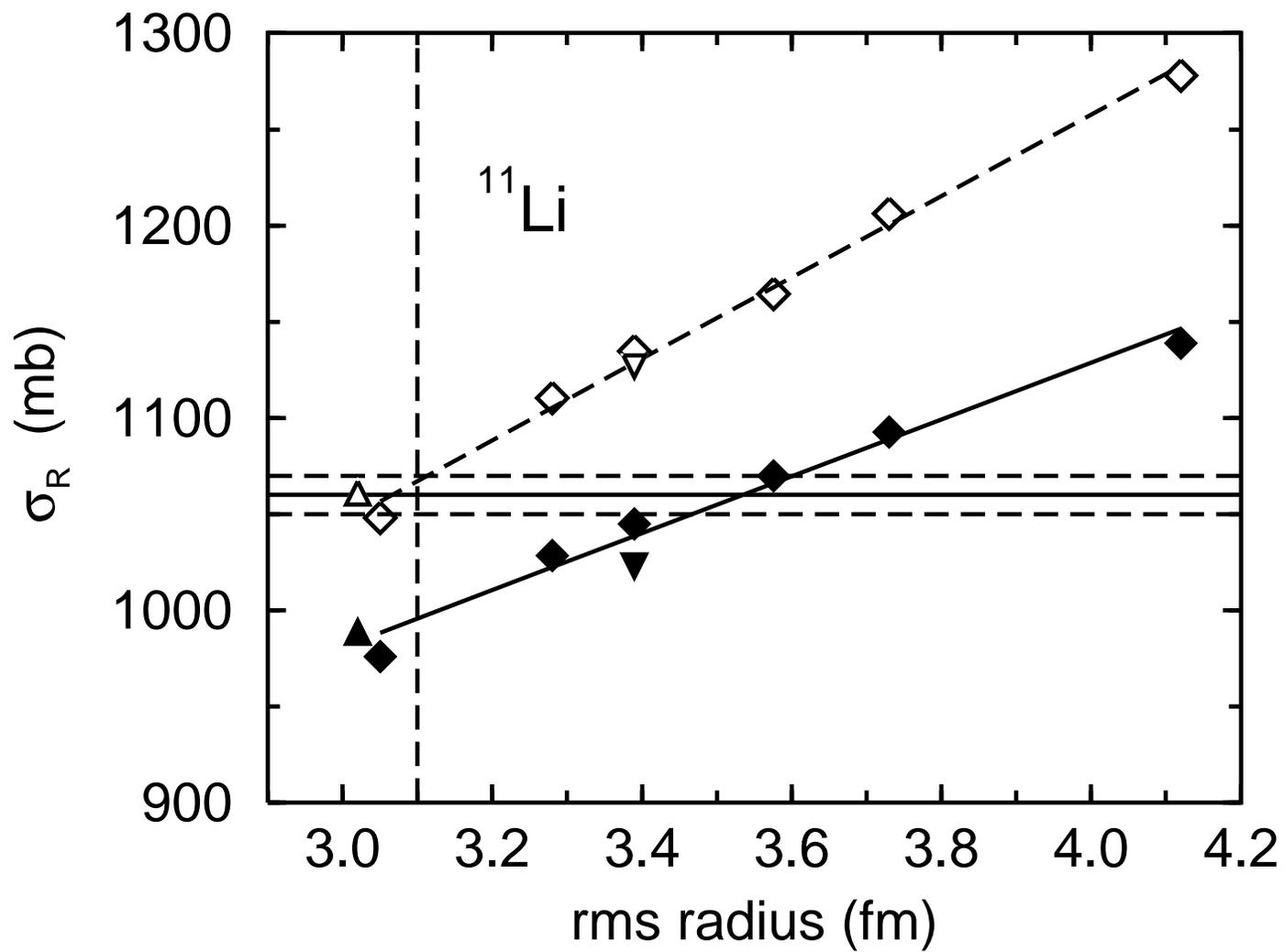
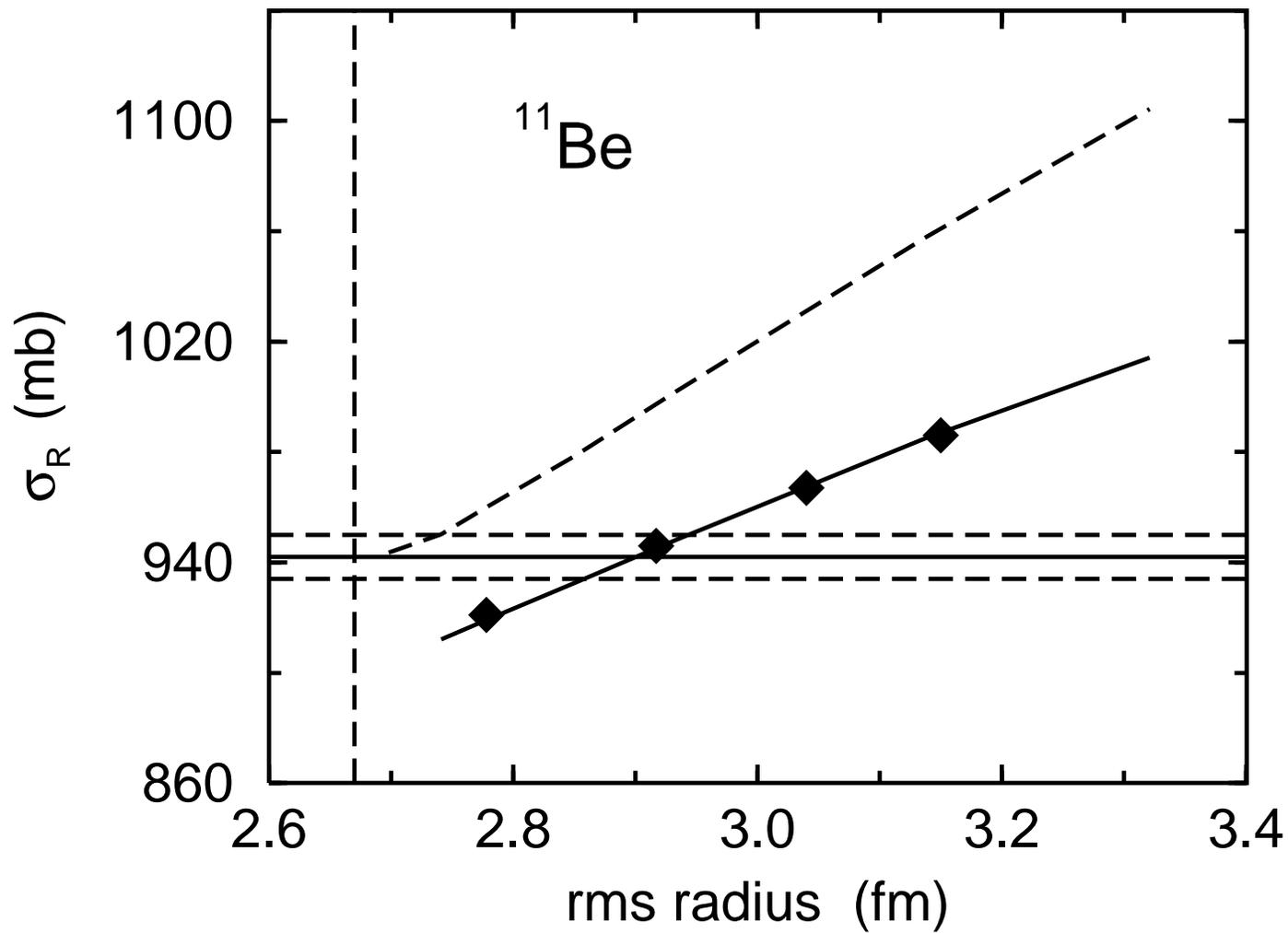
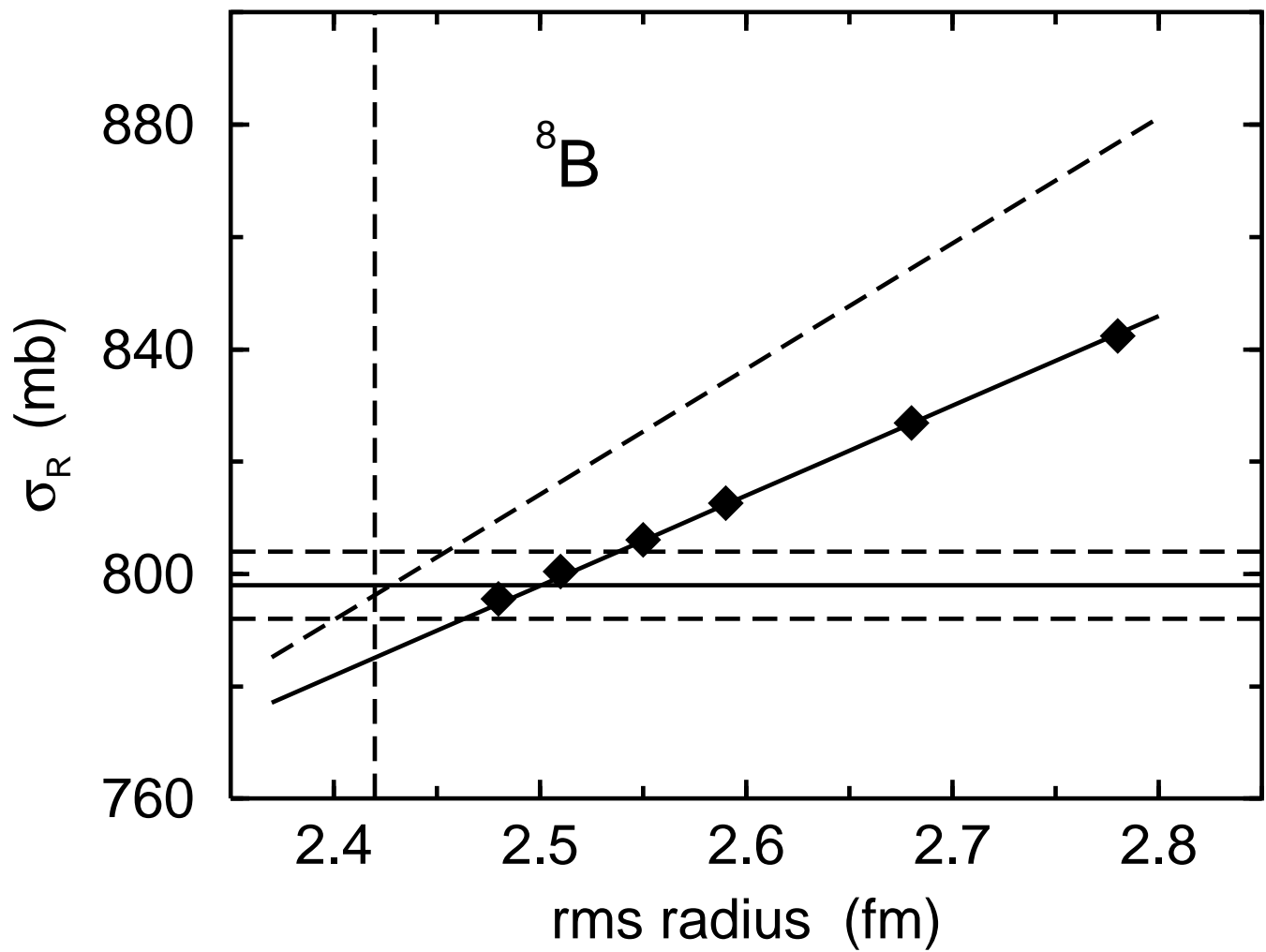
\end{document}